# Mirage: Defense against CrossPath Attacks in Software Defined Networks

Shariq Murtuza
*Department of Computer Science &
Engineering and Information Technology*
*Jaypee Institute of Information Technology*
Noida, India
shariq.murtuza@jiit.ac.in

Krishna Asawa
*Department of Computer Science &
Engineering and Information Technology*
*Jaypee Institute of Information Technology*
Noida, India
krishna.asawa@jiit.ac.in

*Abstract*—The Software-Defined Networks (SDNs) face persis tent threats from various adversaries that attack them using different methods to mount Denial of Service attacks. These attackers have different motives and follow diverse tactics to achieve their nefarious objectives. In this work, we focus on the impact of CrossPath attacks in SDNs and introduce our framework, Mirage, which not only detects but also mitigates this attack. Our framework, Mirage, detects SDN switches that become unreachable due to being under attack, takes proactive measures to prevent Adversarial Path Reconnaissance, and effec tively mitigates CrossPath attacks in SDNs. A CrossPath attack is a form of link flood attack that indirectly attacks the control plane by overwhelming the shared links that connect the data and control planes with data plane traffic. This attack is exclusive to in band SDN, where the data and the control plane, both utilize the same physical links for transmitting and receiving traffic. Our framework, Mirage, prevents attackers from launching adversarial path reconnaissance to identify shared links in a network, thereby thwarting their abuse and preventing this attack. Mirage not only stops adversarial path reconnaissance but also includes features to quickly counter ongoing attacks once detected. Mirage uses path diversity to reroute network packet to prevent timing based measurement. Mirage can also enforce short lived flow table rules to prevent timing attacks. These measures are carefully designed to enhance the security of the SDN environment. Moreover, we share the results of our experiments, which clearly show Mirage's effectiveness in preventing path reconnaissance, detecting CrossPath attacks, and mitigating ongoing threats. Our framework successfully protects the network from these harmful activities, giving valuable insights into SDN security.

## I. Introduction

In the past few decades, computer networks have become increasingly complex, with an increasing demand for more flexible and efficient network management. Traditional net work architecture are often faced with challenges in adapt ing to changing network requirements and ensuring effective communication between devices. To overcome these limita tions, a new paradigm called Software-Defined Networking (SDN) [18] was proposed. Software Defined Networks brings innovative methods for traditional network paradigm. It works by segregating the data plane from the control plane [24, 36], each independent of the other. Usually in traditional networks, the control functions were distributed across various devices, making it difficult to implement network-wide policies and configurations. With SDN, the network now has a software based centralized controller [45], providing much more ef fective management and network control. Along with it the centralization of network orchestration in SDN, enables a network administrator to program and manage the network through software interfaces, making it easier to configure and monitor network devices. This flexibility and programmability offer numerous advantages, including faster deployment of new network services, improved scalability, and enhanced net work visibility. Software Defined Networks allows the network administrators to abstract the underlying network infrastruc ture from the applications running on it. This abstraction [8] enables the network to be more responsive to changing application needs and facilitates the dynamic allocation of network resources. With increasing time, the adoption of SDN has also increased and along with it newer security challenges have also emerged [13]. The presence of a single controller is a big bottleneck in the system [46], since an attack on the controller can render the whole network useless. The control plane is a critical component in managing the network and it is often a potential target of DoS (Denial of Service) and DDoS (Distributed Denial of Service) attacks [40].

A DoS attack [34] is a malicious effort to disrupt a target's regular functioning by overwhelming it with a flood of illegitimate requests or by exploiting vulnerabilities in the system. The aim of the DoS attack is to make a network or service inaccessible to its peers users, resulting in disruption or loss of service [35]. In a DoS campaign usually an attacker overwhelms the target system's resources, such as bandwidth, processing power, memory, or network connections, by flood ing it with an excessive volume of network packets [43]. When these packet originate from multiple hosts under the control of the attacker, it is called as a Distributed Denial of Service (DDoS) attack [39]. Compared to a denial of service attack, a distributed denial of service attack is more complex and difficult to mitigate since source identification is very difficult [6].

These attacks are further classified as follows:
- *Volumetric attacks*: A volumetric attack is targeted to consume all possible network resources such as

band width, CPU resources or even flood a complete network infrastructure. Usually they are performed using UDP floods and ICMP floods [42].
- *Attacks based on protocol vulnerabilities*: These attacks are only possible in networks that are having certain protocols or software deployed, that have a vulnerability which can be exploited, This results in the exhaustion of network resources such as bandwidth, latency or CPU cycles. An often used attack is flooding TCP SYN pack ets. This attack misuses the three-way TCP handshake process to consume server resources [22, 44].
- *Application layer based attacks*: These attacks are also called as layer 7 attacks since they target an application deployed on the network. These attacks usually take form of exploitation of some weakness in the application itself. A HTTP flood is an often used attack under this category [47].

Link Flood Attacks are a category of DDoS attacks that aims to cut connectivity of the network or a sub portion of the network by flooding specific network links [58]. This attack uses very specific tactics to either flood a bottleneck link, or flood the paths carrying incoming and outgoing links to a sub network. Link flooding attacks pose extreme threat to any network infrastructure since they can flood the available network links with an excessive amount of traffic, thus making it impossible for legitimate network traffic to go through these links [59]. These attacks exploit the limited resources of network links, typically bandwidth, causing severe network congestion resulting in degraded performance and high la tency. The primary goal of link flooding attacks is to deplete the network bandwidth, by flooding the links with traffic. An attacker typically achieves this by bombarding the targeted links with a large number of network packets. The attackers usually have a substantial number of decoys nodes that collude with the attacker, helping the attacker in launching attacks from different location across the network.

Just like traditional networks, SDNs are also vulnerable to link flood attacks [55]. SDNs separate the data and the control plane, providing unified network orchestration. A network administrator can monitor and program the SDN accordingly. However, this separation also introduces new security chal lenges, particularly the susceptibility to link flooding attacks. Link flooding attacks exploit the network topologies, aiming to overwhelm network links with an excessive volume of traffic. This flooding saturates the available bandwidth, resulting in disruption to the normal flow of network communication. As a consequence, these attacks negatively impact the performance, reliability, and availability of SDN networks.

The impact of link flooding attacks in an SDN environment can be significant [53]. The reliance on centralized controllers and software-based network management creates a single point of failure, making SDN networks vulnerable to attacks target ing the control plane. Attackers can disrupt communication between switches and the controller by flooding the links, leading to service degradation or even complete network failure.

In this work, we introduce an novel SDN framework Mirage, that utilizes probes to identify and detect ongoing attacks. Our framework also incorporates multiple mechanisms aimed at thwarting all attempts at adversarial network recon naissance, effectively rendering the attacker clueless about existence of shared links. To the best of our knowledge, our work is the first to manipulate flow table rules in order to distort an attacker's perception of the network, thereby preventing CrossPath attack that relies on the attacker gaining knowledge of the shared links.

## II. RELATED WORK

Martin Casado way back in the year 2006 proposed SANE [10] and ETHANE [9] to separate the network into control and data planes for achieving better control over network. With the passage of time, this concept evolved into what we now com monly refer to as Software Defined Networking (SDN) [36]. SDNs logically separate the network flow logic (forwarding plane) from the controlling logic (control plane). They enable programmable control over network routing logic and allow for dynamic changes [1]. In an SDN, the centralized controller not only monitors the network but it also periodically issues flow rules telling switches on which paths to send packets. The switches connected to an SDN are specialized switches with built-in flow tables [33]. The switches are often called as dumb switches since they are unable to make any decision on their own. Upon the arrival of a packet at a switch, the switch checks the flow table to find a matching rule that dictates the action to be taken for that packet, such as forwarding or dropping it. If there is no existing rule that matches, the packet is sent to the controller for further response. The controller shall then provides the appropriate flow table rule. SDNs also allow for variable bandwidth allocation based on application requirements, enabling bandwidth adjustments during heavy loads that can be later restored.

In the last few decades computer networks have faced vari ous types of fierce network attacks from different adversaries, ranging from simple script kiddies to nation-sponsored attack ers [23, 52]. One common method to disrupt communication systems is through Denial of Service (DoS) attacks [12], which aim to stop all incoming and outgoing messages on the targeted victim network host. DoS attacks typically originate from a single source, making their avoidance, detection and mitigation relatively straightforward. Like an arms war, at tackers quickly upgraded their warfare infrastructure to create an upgraded version of DoS attacks called the Distributed Denial of Service (DDoS) attacks [12]. In a DDoS attack, the attackers launch attacks from multiple independent individual sources. These individual sources may have limited attack bandwidth, their combined attack can have devastating effects on the target. The distributed nature of these attacks give them resilience and robustness against detection and mitigation measures that are typically employed by network adminis trators. The selection of individual nodes that send these packets is carefully planned by the attacker after mapping the network in order to maximize the effectiveness of the attack. Consequently, detecting such attacks becomes increasingly challenging [17]. The malicious nature of these attacks lies in their randomized timing, packet routes, and other decisions based on the network's topography. These factors ultimately lead to the disruption of network links. Link Flooding Attacks encompass various variants, with Coremelt [48] and Crossfire [27] attacks being the most widely recognized examples. These attacks utilize diverse methods to disrupt the continuous flow of packets across the network, often by flooding specific links and rendering the targeted victim inaccessible and unavailable. In comparison to traditional Denial of Service (DoS) and Distributed Denial of Service (DDoS) attacks, link flooding attacks require a higher level of sophistication but fewer points of presence across the targeted network to launch an attack. Previous research has explored various approaches, including machine learning and packet inspection, to prevent and mitigate these types of attacks [19].

### III. AN OVERVIEW OF LINK FLOODING ATTACKS

Software Defined Networks have been previously targeted by link flooding attacks [55], with the Coremelt attack and Crossfire attacks being two variants of such attacks. The fundamental principle behind both these attack variants is the same: flooding specific link(s) in the network to isolate the target victim by disrupting incoming and outgoing traffic. This results in the isolation of a single target or maybe even a smaller sub-network of the network. By flooding specific links with a high volume of traffic, the attacker disrupts the normal flow of network packets and can cause significant performance degradation or even network-wide outages [29]. The first variant, the Coremelt attack, has simple requirements [60]. The first and foremost condition is that the targeted network should be be having a dumbbell shaped topology, with nodes on either side on a single critical link that join two different parts. Let the network be divided into two sides, side A and side B. The connection between both sides is dependent on a single link, which plays a critical role but also acts as a bottleneck. All traffic originating from one side and destined for the other side must pass through this lone link, making it a vital pathway with limited capacity. Once the attacker is able to flood this link, the two sides A & B won't be able to communicate with each another due to high latency on the route. To congest the link the attacker places decoys on side A and side B, with the decoys on one side communicating with the decoys on the other side. These pairwise flows lead to traffic flooding on the route and cut off the nodes on side A from all the nodes on side B.

The second variant of the link flooding attack is known as a crossfire attack [27], which builds upon the coremelt attack. This attack focuses on isolating a sub-network of nodes from the main network. In this attack, the sub-network being cut off from the main network does not experience traffic flooding because the attacker strategically places decoys in calculated locations. It is important to note that not all networks are vulnerable to crossfire attacks [2]. Only specific network topologies that include a sub-network with a limited number of paths for incoming and outgoing traffic can be targeted by this particular variant of the attack. In this work, we shift our focus to a recently proposed variant of the link flooding attack known as the CrossPath attack [56]. The CrossPath attack is newly proposed variant of link flood attacks that exploits the knowledge of links shared between data plane and control plane to target critical links. In the following subsections we first discuss what are in Band and out of band SDN then we formally define the CrossPath attacks.

### A. In Band and Out of band SDN

In band [7] and out of band [26] are different network infrastructure design approaches for managing and controlling network traffic. In an in Band configured SDN, the data traffic and control traffic flow through same physical medium. The control channel, which carries control messages from controller to the switch and vice-versa, is established in same physical medium that carries the data traffic. In an In Band SDN deployment, the control channel is typically implemented using the same data forwarding elements (switches) that handle the regular data traffic. This means that the control traffic and data traffic traverse the same network links, uses the same switches, and share the same network resources. The use of in Band SDN offers some advantages, including simplicity and cost effectiveness, as it eliminates the need for a separate control network infrastructure [7]. It simplifies network management and maintenance since control messages and data traffic flow through the same paths. Additionally, it allows far more flexibility in deploying and scaling SDN networks. However the use of an in Band system also exposes control plane packets to competition with data plane packets.

An out of band SDN is link configuration where the data and control traffic are kept separate not only logically but physically as well, using separate physical medium [57]. In an out of band SDN deployment, a separate network infrastructure is used for the control traffic. This can include dedicated communication links, switches, or virtual network overlays that are specifically designed for carrying control messages from switches to the controller and vice-versa. The functioning of the control plane is independent of the data traffic paths, ensuring isolation and dedicated resources for control communication. The use of out of band SDN provides enhanced security and reliability for the control plane. Due to the segregation between data and control planes, susceptibility to interference or attacks that may affect the regular data forwarding is highly reduced. The dedicated control infrastructure allows for better control channel protection, isolation, and monitoring [4].

Out of band SDN can offer higher levels of security and resilience compared to in Band SDN, but it can also introduce additional complexity and cost. The deployment of a sepa rate network infrastructure for the control channel requires additional network elements and configuration. However, in scenarios where security and reliability are paramount, such as critical infrastructure networks or large-scale deployments, out of band SDN may be preferred.

*B. Crosspath attack*

The Crosspath attack [5, 56] is a novel variant of link flooding attacks that can only occur in a network with an in band network design. In other words, it is specific to networks that utilize shared network links for both data and the control plane traffic. A CrossPath flood is a novel threat that targets the control plane in an SDN. Due to the decoupling of data and control plane traffic in SDN, the central controller orchestrates the network by sending messages to the SDN switches via the control plane to exchange network updates. Hence it it imperative to secure the control channel for normal working of the SDN.

The CrossPath attack exploits the existence of shared physi cal links amidst the data and control plane in an SDN [56]. The common links provide an avenue for the attack to disrupt the normal flow of network traffic and potentially cause network instability. The attacker exploits these common links to disturb the control channel. The novelty of this attack lies in its characteristic of not sending any packets on the control plane. Instead, the disruption of the control plane is achieved by the attacker sending specially crafted packets that are transmitted over the data plane. This unique approach allows the attacker to manipulate the network and potentially compromise the control plane's functionality without directly targeting it [5]. These packets interfere with the movement on the control plane. These attack packets remain confined within the data plane and don't enter the control plane, due to which their detection is very difficult. In order to launch this attack, the attacker must initially identify the paths that are shared between the control plane and the data plane. To acquire this information, the adversary must gather details about the physical topology and the paths used for routing within the network.In [56], the authors propose a novel method for identifying shared links. The attacker begins by reconnoitering the network, observing any delays in the control messages that flow through the control plane when a large amount of data plane traffic is sent within a short time window. Through the measurement of control plane message latency in the presence or absence of a brief burst of data plane traffic, the attacker can readily discern which links are common to both the control plane and data plane. This is because only on a shared link can a burst of data plane traffic slow down the control plane traffic. The control traffic can be generated by repeatedly creating new flows, as every flow's first packet is sent to the controller.

The CrossPath attack has significant impacts on various SDN applications. Experimental evaluations have demon strated that it can disrupt crucial services in SDN controllers, resulting in a degradation of performance for applications such as Reactive Routing, Load Balancer, Learning Switch, , and ARP Proxy. These effects manifest as longer response times, decreased reply counts, failure to install forwarding rules, inaccurate routing information, abnormal routing behavior, resetting of flow tables, and excessive load on network links.

If the attacker persists in generating frequent bursts of data traffic, it could eventually lead to the expiration of control plane packets due to timeouts.

IV. MIRAGE: DETECTION OF CROSSPATH ATTACKS

Identifying whether the control plane is currently congested or not can be challenging. Often, network packet drops occur due to factors beyond the control of the administrator. The con trol plane is responsible for executing packet processing tasks such as routing, forwarding, and applying network policies. If errors or failures occur during these processing tasks, packets may be dropped. Such issues can arise from software bugs, hardware failures, or conflicts in the control plane logic. In normal situations, there may be a few sporadic packet drops. However, if there is a significant number of packet drops on a particular path, it raises concerns [20]. To address this, Mirage utilizes probes that traverse the network to identify whether packets are being dropped along a given path.In the event that packets are unable to successfully traverse the intended path despite repeated attempts, Mirage activates its notification mechanism and alerts the administrator. Mirage continuously monitors the network to ensure the reachability of all switches at all times. If a switch becomes unresponsive and the controller is unable to establish connection with it, Mirage can promptly alert the network administrator about a potential ongoing attack.

To detect an attack, the controller consistently sends probes to the switches throughout the network and monitors their responses. If, at any point, a probe fails to receive a reply from a switch, the controller initiates a second round of probes. If the switch still does not respond, it is determined to be under attack, and an immediate alert is sent to the network administrator. The network monitoring probe scheme is presented in detail in the Algorithm 1.

For each switch in the network, Mirage sends a probe message via the control plane and sets the *probeResponseReceived* flag as *f alse*. It then enters a nested loop where it waits for the appropriately chosen *probeInterval* time interval and sends additional probe mes sages to the switch. The number of probe attempts is deter mined by the *maxP robeAttempts* parameter. If, at any point, the *probeResponseReceived* flag remains false, indicating that the switch did not respond to the probe messages, the switch is declared as unreachable. The network administra tor is immediately alerted about the potential attack on the switch's path, following which Mirage's mitigation measures shall be activated.

V. MIRAGE: PREVENTION & MITIGATION OF CROSSPATH ATTACK

In this section we describe the prevention and mitigation features of our framework Mirage. This module empow ers SDN administrators to hinder reconnaissance attempts, specifically adversarial path reconnaissance. By leveraging this module, administrators can effectively prevent Adversarial Path Reconnaissance (APR) attempts from being executed,

**Algorithm 1: Monitoring probes to detect congestion in Control Plane**

1: Initialization: *probeInterval* ← 2 seconds Set *probeInterval* as the time interval between consecutive probes
2: *maxP robeAttempts* ← 2
   Set *maxP robeAttempts* as the maximum number of probe attempts before declaring a switch as non-reachable
3: while Network is operational do
4:   for each switch *S* in the network do
5:     Send probe message to *S* via the control plane
6:     Set probeResponseReceived as false
7:     for *i* = 1 to maxProbeAttempts do
8:       if probeResponseReceived is false then
9:         Wait for probeInterval
10:         Send probe message to *S*
11:       else
12:         Break
13:       end if
14:     end for
15:     if probeResponseReceived is false then
16:       Declare switch *S* as non-reachable
17:       Alert network administrator regarding the possibility of a potential attack on *S*
18:     end if
19:   end for
20: end while

even during the planning stage of an attack. Additionally, if an attacker has already initiated an attack, these same steps enables proactive mitigation measures to minimize the attack impact. This comprehensive approach significantly reduces the likelihood of successful attacks on the SDN infrastructure.

For techniques such as APR to work correctly, timing is the main key [56]. Any mismatch in packets flow calculation can lead to incorrect latency measurement that can render the whole attack fruitless. Our framework Mirage aims to prevent the attacker from making precise measurements thus prevent ing him from gaining network path knowledge. The attacker measures the delay in control plane traffic after sending short data plane traffic bursts. The attacker determines control plane delay by measuring the difference in Round Trip Time (RTT) [28] between the first two packets of a new flow. Once the first packet is sent, control plane messages are triggered, and new flow rules are pushed before receiving the acknowledgement for the second packet. In an ideal network, users should not have the capability to perform traffic monitoring [11]. However, it is challenging to enforce this limitation since we have no control over the hardware and software utilized by the attacking host.

Indeed, we cannot entirely prevent the measurement of Round Trip Time (RTT) between the first two packets of a flow. However, we have the capability to modify or obstruct accurate RTT measurements. This can be accomplished by introducing additional latency in the network, thereby influ encing the timing and measurement of RTT. However an interesting point to note is that merely adding latency to all the packets in all the flows is not going to benefit us, because doing so shall result in the addition of a constant in the original RTT value. Random latency in the measurements of packet is also not beneficial since it may increase failure rate and decrease the Quality of Service (QoS) [37]. To counter an attacker's ability to measure the Round Trip Time (RTT) of the first two packets of a flow, one approach is to prevent them from obtaining this measurement opportunity. Since RTT is calculated at the sender's side and does not rely on the network infrastructure, additional methods are required to manipulate the results and distort the RTT measurements.

*1) Exploiting Path Diversity:* Path diversity refers to the existence of multiple alternative paths between two nodes in a network [21, 50]. It is a characteristic of a network topology where there are multiple routes available for data packets to traverse from a source (S) to a destination (D). For each source-destination (SD) pair, if we have at least two different paths, any attempt by an attacker to identify shared paths can be thwarted. We can assign one path for data traffic and another for control traffic, or alternatively, we can randomly select a path from the available pool of paths. This not only ensures robustness of the network but also allows us to use different paths for data and control channel traffic in case of an ongoing attack [3]. As mentioned in [56], the attacker calculated the time difference between the RTT of the first packet and the second packet of a flow, while simultaneously sending traffic burst, to identify links that are shared between control and data plane traffic. We propose utilizing path diversity as a strategy to avoid detection and prevent such attacks.

The controller will store different paths between same source destination pairs. When the first packet of a flow is received control plane messages will be sent to the con troller and the attacker simultaneously starts data plane traffic burst. However the controller will send two different paths as response to the switch (who initiated the first control plane query). The first path is for the first packet, while the second path (typically a bit less optimal than the first path) is for the second packet. The attacker will get a significant difference in the RTT values $\delta$. This shall lead to high number of false positives. The authors in [56], have mentioned that regardless of the circumstances, certain packets will always be processed solely by the controller. These packets include ARP [15] and DHCP [51], among others. An attacker can exploit these packets to conduct control plane measurements. When an attacker, say h1, is connected to switch s1 and ARP or DHCP traffic is sent, initially, this traffic traverses through h1-s1 link, which is part of the data plane. Once the packet is received by switch s1, the switch forwards the packet to the controller via the control plane. The subsequent path that the packet follows occurs within the control plane. Therefore, by performing straightforward calculations, the user can determine the duration taken by the packet to arrive st the SDN controller C from s1 by subtracting the duration for the packet to reach h1 to s1 from the total time taken. In order to thwart measurement done using these specific techniques we propose the usage of different paths between a source destination pair. For each direction a different path shall be followed if available, that is, while going from h1 to controller path p1 will be followed and the response from the controller will be sent via path p2. This discrepancy in time measurements makes it difficult for the attacker to determine with certainty whether the packet traveled on a shared path or not, thereby hindering their calculations. To identify a probable path out of all available paths between the same source destination, we propose a variant of Equal Cost Multi Path routing (ECMP) Algorithm. The Equal Cost Multi Path is an often deployed routing algorithm in networks to direct network packets from their source to their destination [14, 38]. ECMP (Equal-Cost Multipath) ensures that the selected paths have equal costs. The cost of a path is a parameter that can be adjusted according to the network administrator's specific use case. The algorithm can be based on bandwidth availability, time taken to deliver the packets, hop count etc. ECMP has been widely deployed in traditional networks and SDN as well providing the benefits of load balancing and improved network performance. ECMP has been mainly used to evenly allocate network resources and to maintain similar load on paths having the same source and destination. This leads to highly robust network and fault tolerances [25]. The algorithm selects a probable path based on different criteria such as IP address hash maps, port numbers etc. We now propose our next algorithm, shown in Algorithm 2, for selecting a candidate path from the pool of all possible paths available with same source destination pair.

Our proposed algorithm is a modified version of the original ECMP algorithm. It aims to select the current path with the lowest cost from a range of available paths for directing a flow. Let's go through the algorithm step by step. The algorithm takes as input the flow $F$ and the paths $P_1, P_2, \ldots, P_n$ with their respective costs $C_1, C_2, \ldots, C_n$. For initialization we set the lowest minimum cost (*minCost*) as infinity, the candidate selected path (*selectedPath*) is initialized to null. During each iteration, we go over all available paths starting from $P_1$ to $P_n$. If the cost $C_i$ associated with current path $P_n$ is less than the current minimum cost, the minimum cost variable *minCost* is updated to the cost of the current path, and the selected path *selectedPath* is updated so as to point to the current path $P_i$. The algorithm continues till there are incoming packets from the flow $F$. Once a path has been selected, the algorithm then sends the packet through the selected path *selectedPath* using the *sendPacket(packet)* operation.

Using our variant of ECMP algorithm, all packets belonging to a given flow F will be routed through the currently available path with the lowest cost. Any path that is selected by our algorithm will be removed from the pool of all possible paths, to prevent a single low cost path from being repeatedly being selected. Once all paths have been selected at least once, the

**Algorithm 2:** Modified ECMP Algorithm for Identifying Least Cost Alternate Path Available

1: Input: Packet $p$ from Flow $F$, Paths $P_1, P_2, \ldots, P_n$ with costs $C_1, C_2, \ldots, C_n$
2: Output: Routed packets for flow $F$ via the least cost path
3: Initialization:
4: $minCost \leftarrow \infty$ {Initialize minimum cost to infinity}
5: $selectedPath \leftarrow$ null {Initialize selected path to null}
6: $pathPool \leftarrow P_1, P_2, \ldots, P_n$ {Initialize path pool with all paths}
7: for $i$ = 1 to $n$ do
8: if $C_i < minCost$ then
9: $minCost \leftarrow C_i$ {Update minimum cost}
10: $selectedPath \leftarrow P_i$ {Update selected path}
11: end if
12: end for
13: Routing:
14: $selectedPath$.sendPacket($p$) {Send packet $p$ via the path identified above}
15: $pathPool \leftarrow$ pathPool.remove($selectedPath$) {Remove the path that has been used in this iteration from the pool}
16: if pathPool is empty then
17: $pathPool \leftarrow P_1, P_2, \ldots, P_n$ {if pool is empty, reinitialize the pool with all potential routes from s (source) and d (destination)}
18: end if

pool of available paths will become empty. At this stage, the pool will be replenished by adding all the paths again. At any point throughout the working of the algorithm the controller may updates the costs associated with the paths, add new paths and/or remove old paths. We discuss our findings in the result section.

*2) Short lived flow table rule:* Due to the design of the network infrastructure, path diversity [49] may not be possible always or it may be possible, but only for a few paths. Therefore, we propose our second method, which can be employed when path diversity is not applicable. This method involves using extremely short-lived flow rules. Whenever a new flow starts, the initial packet is passed to the SDN controller. Ideally the controller installs flow table rules in all the switches lying on the identified path and then the first and the rest packets are sent on this path. We propose that for the first packet the flow rules which is installed is unique in nature. The flow rules should be specific to the first packet of each flow, not applicable to the subsequent packets. For the second packet and all subsequent packets, new flow rules can be established to handle their forwarding requirements. We present our second algorithm depicting this scheme, Algorithm 3 below. The network administrator can identify a threshold value $\lambda$ that signifies the number of packets beyond which a single common flow rule would be installed. For example,
if $\lambda = 5$, then for the first 5 packets of every flow, individual unique flow table entries will be pushed by the controller. From the sixth packet onwards, a common entry rule can be created that will be applied to all packets. Over time, these short-lived

---

**Algorithm 3: Short-Lived Flow Rules for Path Establishment with Threshold $\lambda$**

1: Input: New flow $F$ with packets $P_1, P_2, \ldots, P_n$
2: Output: Generation of individual flow table rules for each packet $P_i$
3: $\lambda \leftarrow$ Threshold parameter for establishing common flow rules
4: while New packets $P_i$ of the flow arrive do
5:   if $i \leq \lambda$ then
6:     Send $P_i$ to the controller for path identification
7:     Install unique flow rule corresponding to $P_i$ on the switches lying on the identified path
8:   else
9:     Process $P_i$ as per the existing flow rule, if rule not available yet request for common flow rule for $P_i$ on the switches lying on the identified path
10:   end if
11:   Send $P_i$ on the identified path
12: end while

---

flow table rules will naturally expire and be replaced by new rules. The flow rules established by the controller have a very short timeout duration since they are designed to be used only once for the specific packet they were created for. As a result, they will automatically expire and be replaced by new rules as needed.

## VI. EXPERIMENT DESIGN

We performed multiple simulation using Containernet [41]. We experiment with a total of 261 existing real world topologies taken from the Topology Zoo [31, 32] from around the globe. These topologies reflect real life networks and allow us to understand different scenarios in which an attack may be done [30]. The data in the Topology Zoo is typically provided in a standardized format called the Network Description Language (NDL) [54]. NDL is a textual format that describes the structure and characteristics of a network topology. NDL files contain information about the nodes (network devices), links (connections between nodes), and various attributes associated with them. The format includes details such as node IDs, link capacities, link latencies, geographic locations, and other relevant properties depending on the specific network type. Since these topology maps only list network devices and their associated links, 200 different nodes were added in each topology. Routing information is also not provided in the topology dataset therefore we used the Dijkstra's algorithm [16] to find the shortest path between all source destination pairs. Next we added a controller node in the network that can orchestrate the switches to perform packet switching in the network based on the shortest path, a minimum spanning tree or even non optimum paths. We then performed adversarial path reconnaissance on these networks, and obtained similar result as that mentioned in the original CrossPath paper. We experimented with the topologies, and on exploring them, we found that the majority of the topologies had alternate paths creating path diversity in this topology zoo dataset. Only a handful of topologies had a star or linear shaped network design with little or no path diversity. In [56], the authors have already demonstrated the use of a Protection Rule Activator. This activator ensures that if control plane messages fail to reach the switches, the existing flow rules that allow unrestricted data traffic will eventually time out. Consequently, rules that regulate and prevent data traffic from consuming all the bandwidth, which were previously dormant, will come into effect.

The next module designed in [56] is called the malicious flow locator, implemented in CrossGuard [56]. This module utilizes a divide and conquer approach to identify malicious network flows. It continuously divides the set of flows into two halves and eliminates the half that does not contain the targeted flow.

### A. Mirage path diversity

We will now describe the experimental design of our frame work, Mirage. Our framework utilizes multiple solutions. The first one is path diversity. Normally, network devices construct shortest paths using Dijkstra's Algorithm [16], as depicted in the following algorithm 4 below. To implement path diversity,

---

**Algorithm 4: Dijkstra's Algorithm for Shortest Paths**

Result: Shortest path from a given source vertex $s$ to all other vertices in a graph $G = (V, E)$
Input: Graph $G = (V, E)$, source vertex $s \in V$
Output: Shortest path distances $dist(u)$ and shortest path predecessor nodes $prev(u)$ for all $u \in V \setminus s$
// Initialize $dist(u)$ and $prev(u)$ for all nodes forall $u \in V \setminus s$ do
  $dist(u) := \infty$ $prev(u) := nil$
end
// Initialize the source node $dist(s) := 0$
// Use a priority queue to keep track of the nodes to visit $Q :=$ empty priority queue insert $s$ into $Q$ with priority $dist(s)$
while $Q$ is not empty do
  $u :=$ node in $Q$ with minimum priority remove $u$ from $Q$ forall neighbors $v$ of $u$ do
    $alt := dist(u) + w(u, v)$
    if $alt < dist(v)$ then
      $dist(v) := alt$
      $prev(v) := u$
      insert $v$ into $Q$ with priority $dist(v)$
    end

the controller must have a list of all paths possible between a source destination pair. To identify all possible paths we perform a graph traversal building a list of all possible paths reachable from the current source. This algorithm is shown in Algorithm 5.

**Algorithm 5: Algorithm to Identify All Paths Between All Nodes**

Result: Calculate all possible paths between all source destination pair in a graph $G = (V, E)$
Input: Graph $G = (V, E)$
Output: Path Diversity *paths* between all nodes *paths* := empty list
forall $u \in V$ do
 forall $v \in V$ do
  if $u \neq v$ then
   *path* := empty list *visited* := empty set
   DFS$G, u, v, path, visited, paths$
  end
 end
end
DFS$G, current, end, path, visited, paths$
*visited*.add(*current*) *path*.append(*current*)
if *current = end* then
 *paths*.append(*path*.copy()) Add current path to the list of paths
end
else
 forall neighbor *w* of *current* do
  if *w* not in *visited* then
end
end
end
Recursive DFS call
DFS$G, w, end, path, visited, paths$
*visited*.remove(*current*)
Backtrack by removing current node from visited set *path*.pop()
Remove current node from current path

As can be seen, the controller calculates all possible paths between different source destination pairs and can select different paths between the same source destination pair. Mirage uses path diversity to use alternate paths for to and fro movement of packets between a source destination pair. Mirage thwarts the attacker by sending the control plane requests and responses via different paths. This results in an incorrect RTT calculation, which is required to be highly precise by the attacker in order to mount an attack.

### B. Mirage Short lived flow table rules

The next step taken by mirage is the installation of short lived flow table rules. Mirage ensures, that for the first few packets of a flow, even if only one path is available, the flow table rules are installed on per packet basis. The flow table rules will be created individually for each packet. Effectively this means that for each packet packets of a flow will measure same time taken for the packets to reach destination. The RTT for each packet will contain the time for control plane querying and then time for data plane traffic forwarding.

### VII. POTENTIAL DRAWBACKS OF MIRAGE

In this section, we explore the drawbacks that may arise when deploying our Mirage framework, specifically focusing on attack detection and mitigation steps. Similar to other security measures, the deployment of Mirage in an SDN introduces overhead. The primary overhead experienced by computer networks is network latency, which occurs as packets travel across the network. This latency is typically caused by delays in packet processing at network devices. Likewise, Mirage incurs a cost, not in packet processing, but rather in the identification of paths for packet forwarding, resulting in additional latency.

#### A. Effects of Continuous detection probes

In this section we highlight the issues that we found while deploying continuous monitoring probes. We also present a short discussion on how can these issues be tackled.

*1) False Positive:* If the probe responses are unable to reach back to the controller, either due to delay in processing or due to some other non malicious reason, the system may still trigger an alarm stating that a switch is non reachable due to an attack. This false positive can be verified manually by the administrator by tracing the path to the switch.

*2) Control Plane Traffic:* Mirage periodically sends out probes across the network to check the connectivity with the SDN switches. This constant monitoring may give rise to a small network overhead. This overhead grows proportionally the switch will ask the controller for instructions, hence any attacker waiting for the opportunity to measure time difference between consequent with the rise in the number of SDN switches. This can cause stability issues in extremely large networks with thousands of switches. To minimise these side effects we propose the following methods.

1) Tuning of *probeInterval* and *maxP robeAttempts* provides strong a control on the flux of packets across the network.
2) Instead of sending packets simultaneously, Mirage can be configured to send probes in batches using a round robin configuration. Each batch of probes will be initi ated only once the previous batch has been completely processed.

Both these adjustments can ensure that the control plane is not flooded by the monitoring probes themselves.

#### B. Effects of Our Mitigation Strategies on the Normal Func tioning of the SDN

Mirage has two mitigation modules, the first module uses ECMP variant to identify paths for different packets of the same flow. The second module uses short lived flow rule for network management packets (ARP, DHCP, etc.) so that each packet may invoke unique control plane query. Both the modules hinder adversarial path reconnaissance attempts and, instead, produce inaccurate measurements if an attacker calculates the round trip time (RTT) difference of first and second packets of a flow.

*1) Effects of Alternate Routing in Mirage :* The first module employs a variant of ECMP (Equal-Cost Multipath) routing to determine the most suitable path for a packet $P_i$ belonging to the flow $F$ from a pool of available paths. When a path is chosen from this pool, it is removed and will not be selected again until all paths in the pool have been exhausted. Once the entire pool of paths has been exhausted, all possible paths between source-destination pairs are added back into the pool.

- The first overhead incurred by this scheme is that the controller must store all possible paths between each source-destination pair, along with their associated costs, and update them frequently. Previously, the controller was required to only store the shortest path between a source and destination, along with a backup route that would be used in emergency situations.
- The second overhead manifests itself as additional control plane traffic generated during the querying and response process in Algorithm 2. For each packet $P_i$ in flow $F$, we need to follow the same procedure of requesting a new path from the controller. In response, the controller pushes the relevant rules to the switches. However, due to the short lifespan and quick expiration of flow rules, they are rapidly replaced by newer rules. This process leads to a significant influx of control plane traffic in the network, occurring at an approximate 1:1 ratio. In other words, for every new packet $P_i$ in flow $F$, a new control plane request is generated. This drawback of using Mirage results in increased overhead in control plane traffic, which is necessary to prevent adversarial path reconnaissance.

*2) Effects of Mirage Short Lived Flow Rules:* The second module is responsible for installing short-lived, unique flows that correspond to individual packets of flow $F$ when the previously discussed alternate routing cannot be deployed due to the unavailability of alternate paths or in the case of packets belonging to DHCP, ARP etc. In this module, for the first $\lambda$ number of packets in flow $F$, each packet queries the controller. This prevents the attacker from accurately measuring the Round Trip Time (RTT) between the packets because they lack knowledge of which packet's RTT includes the control plane querying time. The network administrator can select the variable threshold, $\lambda$, which represents the number of packets. The overhead incurred by this module involves generating additional control plane traffic by querying the controller for the flow rules for the first $\lambda$ number of packets.

## VIII. RESULTS

We deployed 261 real-world topologies based on the Topology Zoo dataset, upon which we conducted our experiments and corresponding measurements. Below, we discuss our findings. The Topology Zoo dataset consists of a total of 261 different network topologies, including various configurations such as ring, star, linear, mesh, hybrid, and more. These topologies are derived from real-world network infrastructures, ranging from small-scale experimental setups to large-scale deployments in critical systems like autonomous systems, ISPs, and WANs. As these topologies carry real-life network traffic and are considered critical, we used them as the base topology for our different simulations.

### A. Attack Detection Module

To detect unresponsive switches we deploy probes across the network, via the control plane. In our experiments, for all the toplogies simulated, we introduced congestion in the links that were share by flooding data plane. Simultaneously the control plane was also choked. We then launched Mirage's attack detection probes. Whenever a switch was unreachable we were able to detect it with an accuracy of 100%, which means we never got a false negative. However when the network was under heavy load naturally, in few simulation the replies generated in response to Mirage's probes were unable to reach their destination, resulting in false positives being generated. Table I shows our findings and observations during our experiments to detect unresponsive switches.

TABLE I: Results of Switch Unresponsiveness Detection

| Measurement | Value |
| --- | --- |
| Total number of switches tested | 50 |
| Number of switches detected as unresponsive | 8 |
| True Positives (correctly identified) | 7 |
| False Positives (incorrectly identified) | 2 |
| False Negatives (incorrectly not identified) | 1 |
| True Negatives (correctly identified) | 40 |
| Detection Accuracy | 88% |
| False Positive Rate | 4% |
| False Negative Rate | 12% |
| Mean Response Time for Unresponsive Switches | 529 ms |
| Mean Response Time for Responsive Switches | 154 ms |

Observations and Insights:
- False positives were encountered in a few cases, potentially due to heavy network load during testing.
- The mean response time for unresponsive switches was significantly higher than that for responsive switches.

Variations and Scenarios:
- Experiment conducted under different network loads (low, medium, high): Accuracy rates varied from 90% to 85%.
- Varying network topologies (ring, star, mesh): Accuracy rates showed very minor variations across topologies. Figure 1 shows how network load affects the accuracy. The reduction in accuracy is due to loss of probe packets due to network congestion.

## B. Mitigation using Alternate Routing

Mirage maintains a large pool of available paths between a source destination pair, which are selected on a per packet basis when a new flow is established between that source destination pair. The success of this module is directly dependent on the existence of alternate paths. The topologies available in the topology zoo dataset consist of worldwide available public network topologies. We calculated how many topologies had alternate paths between different source and destination pairs and we found that only 30 topologies had zero redundant paths out of the 261 total topologies.

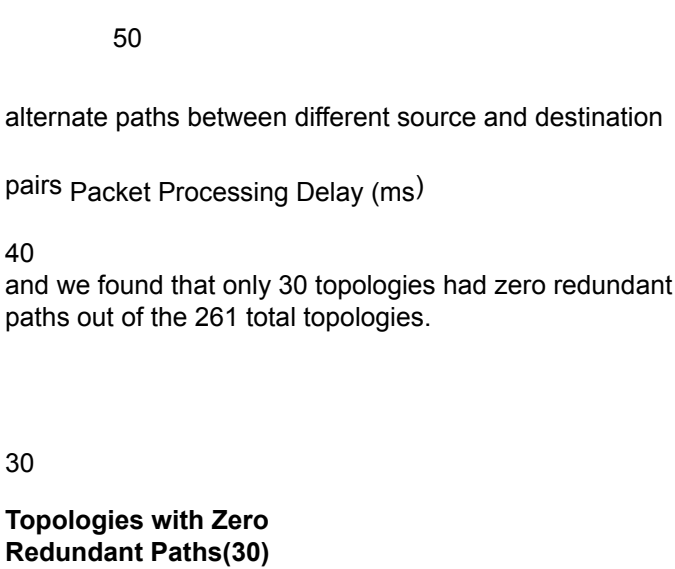

Fig. 1: Relationship between Network Load and Loss of Probe Packets

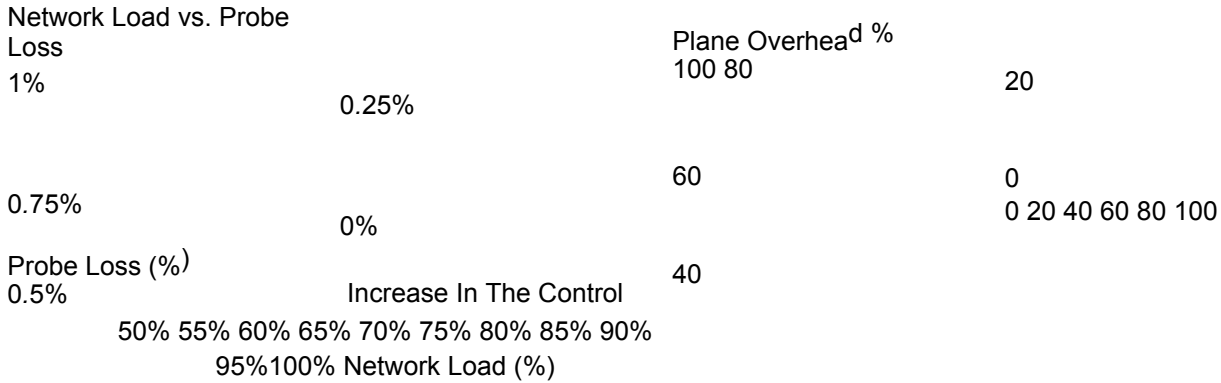

Fig. 2: Existence of Alternate Paths in the Topology Zoo Dataset

Fig. 3: Relationship between Control Plane Load and Flow table Rules

Table III, in the appendix, shows the percentage of source destination pairs in a topology having alternate paths between them. The heatmaps shown in the appendix show the amount of alternate paths between the same source destination pairs.

## C. Short Lived Flows

The short lived flows created by Mirage to thwart reconnaissance attempts often results in an increase in the number control plane packets. This increased packet count resulted in some quality degradation. The control plane traffic load increased with the increase in the number of flow table rules as shown in figure 3. In figure 4 we show the increase on the switch load due to increase in the control plane traffic. Table II shows the overhead incurred by our controller while implementing Mirage's short lived flow rules.

Fig. 4: Increase in Packet processing due to control Plane load.

TABLE II: Controller Overhead

| Metric | Without Mirage | With Mirage |
|---|---|---|
| CPU utilization | 72% | 84% |
| Memory usage | 800 MB | 1017 MB |

## IX. Conclusion

In this work, we presented our framework Mirage, which aims to prevent adversarial path reconnaissance and CrossPath denial of service attacks. Mirage detects the control plane attacks by persistently probing the SDN switches while moni toring their responses. Whenever a response is skipped or not received, additional probes are launched to check the status of the switch, and raise an alert in case of no response even after repeated attempts. Mirage implements preventive measures to thwart adversarial path reconnaissance, which is the initial step in launching a CrossPath attack, thereby preventing the attack. To thwart the reconnaissances Mirage utilizes similar cost paths to route different packets of the same flow, whenever such paths are available. This causes an attacker, who is measuring delays to identify shared paths, to get inconsistent results for the packets. When such paths are not available due to network constraints, Mirage resorts to mandatory querying the controller for each packet of a flow until a specified number of packets, denoted by the threshold $\lambda$, have been processed. After processing these packets, a common flow table rule can be used for the remaining packets, thwarting the attacker's calculations. The overheads associated with Mirage are acceptable, compared to the protection it provides from CrossPath attacks.

Geoffrey M Voelker. In search of path diversity in isp
The below table list some of the toplogies along with the percentage of source destination pairs that have path diversity between them.

TABLE III: Percentage of Paths in the Topology having Path Diversity

| Topology Name | Percentage of Paths have alternates |
|---|---|
| Abilene.gml | 100.00% |
| Arpanet19719.gml | 100.00% |
| Arpanet19728.gml | 100.00% |
| Attmpls.gml | 100.00% |
| Belnet2007.gml | 100.00% |
| Belnet2008.gml | 100.00% |
| Belnet2009.gml | 100.00% |
| Belnet2010.gml | 100.00% |
| Darkstrand.gml | 100.00% |
| Digex.gml | 100.00% |
| Elibackbone.gml | 100.00% |
| Epoch.gml | 100.00% |
| Globalcenter.gml | 100.00% |
| Gridnet.gml | 100.00% |
| Heanet.gml | 100.00% |
| Janetbackbone.gml | 100.00% |
| Netrail.gml | 100.00% |
| Oxford.gml | 100.00% |
| Sanren.gml | 100.00% |
| Vtlwavenet2008.gml | 100.00% |
| Vtlwavenet2011.gml | 99.76% |
| Arpanet19723.gml | 99.67% |
| Surfnet.gml | 99.67% |
| Abvt.gml | 99.60% |
| Missouri.gml | 99.50% |
| Networkusa.gml | 99.50% |
| Quest.gml | 99.47% |
| Internetmci.gml | 99.42% |
| Rediris.gml | 99.42% |
| Ans.gml | 99.35% |
| Hibernianireland.gml | 99.35% |
| Ibm.gml | 99.35% |
| Renater2010.gml | 99.34% |
| Dfn.gml | 99.33% |
| Geant2009.gml | 99.29% |
| Xeex.gml | 99.28% |
| Palmetto.gml | 99.19% |
| Geant2010.gml | 99.10% |
| Funet.gml | 99.08% |
| Atmnet.gml | 99.05% |
| Columbus.gml | 99.05% |
| Biznet.gml | 99.01% |
| Pionierl1.gml | 98.89% |
| Geant2012.gml | 98.85% |
| Savvis.gml | 98.83% |
| Uninet.gml | 98.72% |
| Iowastatewidefibermap.gml | 98.67% |
| Iris.gml | 98.67% |
| Bellcanada.gml | 98.58% |
| Packetexchange.gml | 98.57% |

| File | % |
|---|---|
| Renater2006.gml | 98.48% |
| Renater2008.gml | 98.48% |
| Janetlense.gml | 98.42% |
| York.gml | 98.42% |
| Lambdanet.gml | 98.37% |
| Hostwayinternational.gml | 98.33% |
| Integra.gml | 98.29% |
| Sprint.gml | 98.18% |
| Gtspoland.gml | 98.11% |
| Hiberniauk.gml | 98.10% |
| Iij.gml | 98.05% |
| Highwinds.gml | 98.04% |
| Intranetwork.gml | 97.98% |
| Switch.gml | 97.82% |
| Belnet2005.gml | 97.63% |
| Belnet2006.gml | 97.63% |
| Pionierl3.gml | 97.58% |
| Hiberniacanada.gml | 97.44% |
| Goodnet.gml | 97.06% |
| Hurricaneelectric.gml | 96.74% |
| Noel.gml | 96.49% |
| Beyondthenetwork.gml | 96.37% |
| Agis.gml | 96.33% |
| Geant2001.gml | 96.30% |
| Canerie.gml | 96.17% |
| Nsfnet.gml | 96.15% |
| Redbestel.gml | 95.96% |
| Aarnet.gml | 95.91% |
| Peer1.gml | 95.83% |
| Cernet.gml | 95.73% |
| Belnet2003.gml | 95.65% |
| Belnet2004.gml | 95.65% |
| Garr200909.gml | 95.56% |
| Garr201005.gml | 95.56% |
| Garr201007.gml | 95.56% |
| Garr201008.gml | 95.56% |
| Gtsczechrepublic.gml | 95.56% |
| Garr200902.gml | 95.53% |
| Garr200908.gml | 95.53% |
| Garr200912.gml | 95.53% |
| Garr201001.gml | 95.53% |
| Garr201003.gml | 95.53% |
| Garr201004.gml | 95.53% |
| Garr201010.gml | 95.52% |
| Garr201012.gml | 95.52% |
| Garr201101.gml | 95.52% |
| Switchl3.gml | 95.47% |
| Easynet.gml | 95.32% |
| Garr201102.gml | 95.30% |
| Arr201108.gml | 95.15% |
| Garr201104.gml | 95.15% |
| Garr201105.gml | 95.15% |
| Garr201107.gml | 95.15% |
| Garr201108.gml | 95.15% |
| Garr201109.gml | 95.03% |
| Garr201110.gml | 95.03% |
| Garr201103.gml | 94.98% |
| Bteurope.gml | 94.93% |
| Nextgen.gml | 94.85% |
| Btasiapac.gml | 94.74% |
| Esnet.gml | 94.73% |
| Garr201111.gml | 94.63% |
| Garr201112.gml | 94.54% |
| Garr201201.gml | 94.54% |
| Arpanet19706.gml | 94.44% |
| Internode.gml | 94.27% |
| Psinet.gml | 94.20% |
| Cwix.gml | 94.13% |
| Aconet.gml | 94.07% |
| Evolink.gml | 93.99% |
| Chinanet.gml | 93.73% |

Shentel.gml 93.65% Ussignal.gml 93.65% Ntelos.gml 93.62%
Bsoneteurope.gml 93.46%
Compuserve.gml 93.41%
Dataxchange.gml 93.33%
Sanet.gml 93.24% Bellsouth.gml 93.18% Niif.gml 92.86%
Bbnplanet.gml 92.59% Airtel.gml 92.50% Rhnet.gml 92.50%
Claranet.gml 92.38%
Hiberniaus.gml 92.21% Karen.gml 91.67% Spiralight.gml 91.43%
Grnet.gml 91.14%
Cesnet201006.gml 90.80%
Bandcon.gml 90.48% Getnet.gml 90.48% Sinet.gml 90.30%
Renater2001.gml 89.49%
Restena.gml 89.47% Bren.gml 89.19% Cesnet200304.gml 89.16% Roedunetfibre.gml 89.01% Unic.gml 89.00%
Cesnet200511.gml 88.66%
Cesnet200603.gml 88.66%
Rnp.gml 88.60%
Asnetam.gml 88.32% Marwan.gml 87.50% Cesnet200706.gml 87.32% Pacificwave.gml 86.93% Iinet.gml 86.88% Fatman.gml 86.76% Kentmanjul2005.gml 85.83% Eunetworks.gml 85.71%
Hiberniaireland.gml 85.71%
Widejpn.gml 84.83% Navigata.gml 84.62% Ernet.gml 83.91%
Arpanet196912.gml 83.33%
Gtsromania.gml 82.86%
Harnet.gml 82.38% Litnet.gml 80.29% Pern.gml 80.08%
Layer42.gml 80.00% Marnet.gml 80.00% Napnet.gml 80.00%
Telecomserbia.gml 80.00%
Vinaren.gml 79.67%
Garr199901.gml 79.17%
Oteglobe.gml 78.49%
Garr200404.gml 77.92%
Garr199904.gml 77.87%
Garr199905.gml 77.87%
Garr200112.gml 76.81%
Garr200109.gml 76.62%
Kentmanaug2005.gml 75.93%
Garr200212.gml 75.78%
Myren.gml 75.53%
Gtsslovakia.gml 72.77% Tlex.gml 71.21% Istar.gml 68.77%
Latnet.gml 67.52%
Btlatinamerica.gml 67.45%
Fccn.gml 67.19% Gtshungary.gml 62.76% Nsfcnet.gml 62.22%
Eenet.gml 60.26%
Deutschetelekom.gml 59.51%
Kentmanapr2007.gml 58.10%
Kentmanjan2011.gml 57.61%
Nordu2005.gml 55.56% Cudi.gml

53.96% Dialtelecomcz.gml 50.77% Twaren.gml 50.53% Gambia.gml 48.94% Cesnet2001.gml 44.27% Ulaknet.gml 41.82% Ilan.gml 39.56% Forthnet.gml 38.02%

APPENDIX

Heatmaps depicting the presence of alternate paths in the topologies are shown in Figure 5. The darker the color of the cell, the greater the number of alternate paths that exist between the source and destination (row, column) pair.

Nordu2010.gml 30.72% Uran.gml 22.83% Ai3.gml 0.00% Amres.gml 0.00% Arn.gml 0.00% Azrena.gml 0.00% Basnet.gml 0.00% Carnet.gml 0.00% Cesnet1993.gml 0.00% Cesnet1997.gml 0.00% Cesnet1999.gml 0.00% Cynet.gml 0.00% Gblnet.gml 0.00% Grena.gml 0.00% Itnet.gml 0.00% Janetexternal.gml 0.00% Jgn2plus.gml 0.00% Kreonet.gml 0.00% Mren.gml 0.00% Nordu1989.gml 0.00% Nordu1997.gml 0.00% Padi.gml 0.00% Renam.gml 0.00% Renater1999.gml 0.00% Reuna.gml 0.00% Sago.gml 0.00% Singaren.gml 0.00% Telcove.gml 0.00% Visionnet.gml 0.00% Zamren.gml 0.00%

Fig. 5: Heatmaps showing alternate paths